\documentclass[universe,article,submit,pdflatex,oneauthor]{Definitions/mdpi} 

\firstpage{1} 
\pubvolume{1}
\issuenum{1}
\articlenumber{0}
\pubyear{2025}
\copyrightyear{2025}
\datereceived{ } 
\daterevised{ } 
\dateaccepted{ } 
\datepublished{ } 
\hreflink{https://doi.org/} 

\usepackage{varwidth}
\providecommand{\tabularnewline}{\\}

\Title{The magnetic helicity driven solar-type dynamo}

\TitleCitation{The magnetic helicity driven dynamo}

\Author{Valery V. Pipin $^{1}$\orcidA{}}

\AuthorNames{Valery V. Pipin}

\address{%
$^{1}$ \quad Institute for Solar-Terrrestrial Physics, Irkutsk, Russia; pip@iszf.irk.ru}

\corres{Correspondence: pip@iszf.irk.ru}

\abstract{
(1)The previous theoretical studies showed that the large-scale vorticity generate  the divergent-type
helicity flux from the magnetic fluctuations. Similarly to the $\alpha$ effect, this  breaks the equatorial reflection symmetry of the magnetic
fluctuations in the stellar convection zone. This effect was called
the new Visniac flux (hereafter the NV flux).  2)Methods. Using the mean-field dynamo model we study the effect of the
NV flux on the solar type dynamos. 3)Results. We found that the NV flux results
to a increase of the dynamo efficiency for the turbulent generation
of the large-scale poloidal magnetic field of the Sun. The dynamic
effect of the NV flux on the magnetic field evolution results into
concentrating the dynamo waves toward the equator. Using the numerical
simulations of the mean-field dynamo model we compare the helicity
production rates by the turbulent dynamo effects, like the $\alpha$
effect and the NV flux. The model shows that the new dynamo source
due to large-scale vorticity and small-scale dynamo results into amplification
of the poloidal magnetic field generation in polar regions at the
top of dynamo domain. Therefore the fluctuating magnetic activity
at high latitudes of the Sun provides the additional source of the
large-scale poloidal magnetic field.
}


\begin{document}

\section{Introduction}

The basic scenario of the solar dynamo was suggested by Parker \cite*{Parker1955a,Parker1955b}.
It is employed in the mean-field dynamo models which simulate the
large-scale dynamo as a cyclic transformation of the poloidal magnetic
field to the toroidal by means of the differential rotation and regeneration
of the poloidal magnetic field from the toroidal by means of the cyclonic
turbulent convection, the $\alpha$ effect, and the systematic tilt
magnetic field inside bipolar solar active regions, i.e., the Joy's
law. The origin of the $\alpha$ effect is related with the effect
of the Coriolis force on the convection in the stratified rotating
convection zone. The origin of the tilt of the bipolar active regions
(hereafter, BMR) is still under debate. Both the $\alpha$ effect
and the systematic tilt of the BMR are due to the breaking of the
reflectional symmetry of the turbulent motions and magnetic field
inside the convection zone. It is noteworthy that the large-scale
dynamo produces the helical large-scale magnetic field with the positive
helicity in the northern and the negative one in the southern hemisphere.
The conservation of the total magnetic helicity results into opposite
hemispheric helicity rule for the small-scale magnetic field. The
theoretical arguments of \citet{Pouquet1975,Frisch1975,Kleeorin1982}
showed that the obtained distribution of the small-scale magnetic
helicity opposes the turbulent generation of the large-scale poloidal
magnetic field in stellar convective zones. 

The nonlinear feedback of the magnetic helicity on the turbulent dynamo
gave a motivation to study the effects of the magnetic helicity transports
to overcome the restrictions of the magnetic helicity conservations
in the large-scale dynamo \citep{Kleeorin2000,Vishniac2001,BrandVish2025ApJ}.
In particular, the results of \citet{Pipin2008a,Kleeorin2022,Subramanian2023ApJ}
(hereafter, P08, KR22, and GS23, respectively) showed that in the
large-scale vorticity can generate the magnetic helicity
even when the original turbulent motions and magnetic fields have
no helicity at all. For this effect the small-scale magnetic fields
that stem from the small-scale dynamo are essential. The small-scale
dynamo is expected to exist in the turbulent media of the stellar
convective envelopes \citep{Kapylae2023SSRv}. In the stellar
convection zone the large-scale vorticity corresponds to the presence
of the differential rotation and the meridional circulation. Both types of the
large-scale flow contribute to generation of the magnetic helicity
flux which is accompanied to the magnetic activity on the solar surface.
The differential rotation is considered as the main source of the
magnetic helicity transport from the solar convection zone to the
outer layers of the Sun \citep{Berger2000,PariDemBer2005AA,Sun2024AA}.
The results of \citet{Prior2014ApJ} and \citet{Pipin2025ApJ} showed
that the differential rotation can be the main source of the hemispheric
helicity rule of the magnetic field in solar active regions. 

Following the results of the analytical studies of \citet{Subramanian2023ApJ}
(hereafter GS23) and results of the above mentioned numerical simulations
we conclude that the differential rotation can be an important source
of the small-scale magnetic helicity in the solar convection zones. Interesting
that this source of the magnetic helicity is rarely included in the
dynamo models. The results of \citet{Guerrero2010} showed that the
differential rotation can be important for the helicity flux. The
so called Visniac-Cho flux \citep{Vishniac2001} which they studied
is nonlinear about the large-scale magnetic field. Therefore, it is
likely important for the supercritical dynamo regimes. However, the
results of P08, KR22 and GS23 showed that the large-scale vorticity
produces the magnetic helicity even for the linear case about the strength
of the large-scale magnetic field. This term should be dominant for
a slightly overcritical dynamos. Here, we study this effect in the
mean-field dynamo for the first time. Our plan is as follows. Next
chapter formulates the dynamo model and the basic assumptions behind
it. In the chapter 3 we consider the results of the dynamo model and
compare the new effects of the helicity generation with the standard
effects of the large-scale dynamo. Finally, in chapter 4, we discuss
the findings and draw the main conclusions of the paper.

\section{The dynamo model}

The evolution of the large-scale magnetic field, $\left\langle \boldsymbol{B}\right\rangle $
, in the highly conductive turbulent media is governed by the induction
equation,
\begin{equation}
\frac{\partial\left\langle \boldsymbol{B}\right\rangle }{\partial t}=\nabla\times\left(\boldsymbol{\mathcal{E}}+\left\langle \boldsymbol{U}\right\rangle \times\left\langle \boldsymbol{B}\right\rangle \right),\label{eq:MFE}
\end{equation}
where $\boldsymbol{\mathcal{E}}=\left\langle \boldsymbol{u}\times\boldsymbol{b}\right\rangle $
is the mean electromotive force of the turbulent flows and magnetic
field; $\left\langle \boldsymbol{U}\right\rangle $ is the large-scale
flow. Here, the angle brackets mark the averaging over the ensemble
of fluctuations. Usually, see e.g., \citet{Pipin2025ApJ} we distinguish
the axisymmetric and nonaxisymmetric components of the magnetic field
and flows:
\begin{eqnarray}
\left\langle \boldsymbol{B}\right\rangle  & = & \overline{\left\langle \boldsymbol{B}\right\rangle }+\left\langle \tilde{\boldsymbol{B}}\right\rangle ,\\
\left\langle \boldsymbol{U}\right\rangle  & = & \overline{\left\langle \boldsymbol{U}\right\rangle }+\left\langle \tilde{\boldsymbol{U}}\right\rangle ,
\end{eqnarray}
where the overline marks the longitudinal average. In this paper,
we study the axisymmetric mean-field dynamo and put $\overline{\left\langle \boldsymbol{B}\right\rangle }=\overline{\boldsymbol{B}}$,
$\overline{\left\langle \boldsymbol{U}\right\rangle }=\overline{\boldsymbol{U}}$.
In following the line of \citet{Krause1980}, the axisymmetric field
is decomposed into the sum of the poloidal and toroidal parts: 
\begin{eqnarray}
\overline{\boldsymbol{B}} & = & B\boldsymbol{e}_{\phi}+\nabla\times\left(\frac{A\boldsymbol{e}_{\phi}}{r\sin\theta}\right),\label{eq:ax}\\
\overline{\boldsymbol{U}} & = & r\sin\theta\Omega\boldsymbol{e}_{\phi}+\overline{\boldsymbol{U}}^{M}
\end{eqnarray}
where the scalars $A$, $B$ and the angular velocity $\Omega$ are
the functions of time and spatial variables: $r$ is the radius and $\theta$ is the co-latitude
(the polar angle);$\boldsymbol{e}_{\phi}$ is the unit vector along
the azimuth; $\overline{\boldsymbol{U}}^{M}$is the meridional circulation.
The details of our model can be found in \cite{PKT23}.
The dynamo model employs the mean electromotive force, $\boldsymbol{\mathcal{E}}$,
as follows, 
\begin{equation}
\mathcal{E}_{i}=\left(\alpha_{ij}+\gamma_{ij}\right)\overline{B}_{j}-\eta_{ijk}\nabla_{j}\overline{B}{}_{k},\label{eq:emf}
\end{equation}
where $\alpha_{ij}$ describes the turbulent generation by the hydrodynamic
and magnetic helicity, $\gamma_{ij}$ is the turbulent pumping and
$\eta_{ijk}$ - the eddy magnetic diffusivity tensor. The $\alpha$-effect
tensor includes the effect of the magnetic helicity conservation (\citealp{Kleeorin1982,Kleeorin1999}),
\begin{eqnarray}
\alpha_{ij} & = & \psi_{\alpha}(\beta)C_{\alpha}\alpha_{ij}^{{\rm K}}u_{c}+\alpha_{ij}^{{\rm M}}\psi_{\alpha}(\beta)\frac{\left\langle \boldsymbol{a\cdot}\boldsymbol{b}\right\rangle \tau_{c}}{4\pi\overline{\rho}\ell_{c}^{2}}.\label{alp2d}
\end{eqnarray}
Here $C_{\alpha}$ is the dynamo parameter characterizing the magnitude
of the hydrodynamic $\alpha$-effect, and $\alpha_{ij}^{{\rm K}}$
and $\alpha_{ij}^{{\rm M}}$ are dimensionless and describe the anisotropic
properties of the kinetic and magnetic $\alpha$-effect. Their radial
profiles depend on the equipartition parameter, $\epsilon=\left\langle b^{(0)2}\right\rangle /4\pi\overline{\rho}\left\langle u^{(0)2}\right\rangle \le1$,
the mean density stratification and the spatial profiles of the convective
velocity $u_{c}=\sqrt{\left\langle u^{(0)2}\right\rangle }$, and
on the Coriolis number, 
\begin{equation}
\Omega^{\star}=2\Omega_{0}\tau_{c},\label{eq_M8}
\end{equation}
where $\Omega_{0}$ is the global angular velocity of the star and
$\tau_{c}$, $\ell_{c}$ are the convective turnover time and the
mixing length respectively. The magnetic quenching function $\psi_{\alpha}(\beta)$
depends on the parameter $\beta=|\overline{\boldsymbol{B}}|/\sqrt{4\pi\overline{\rho}u_{c}^{2}}$
\citep{PK19}. We put the description of $\alpha_{ij}^{{\rm K}}$
and $\alpha_{ij}^{{\rm M}}$ as well as turbulent pumping tensor,
$\gamma_{ij}$ and the eddy magnetic diffusivity tensor, $\eta_{ijk}$.
in Appendix.

To calculate the reference profiles of mean thermodynamic parameters,
such as entropy, density, temperature and the convective turnover
time, $\tau_{c}$, we use the MESA model \citep{Paxton2011,Paxton2013}.
We use the mixing-length approximation to define the profiles of the
eddy heat conductivity, $\chi_{T}$, eddy viscosity, $\nu_{T}$, eddy
diffusivity, $\eta_{T}$, and the RMS convective velocity, $u_{c}$,
as follows, 
\begin{eqnarray}
\chi_{T} & = & \frac{\ell_{c}^{2}}{6}\sqrt{-\frac{g}{2c_{p}}\frac{\partial\overline{s}}{\partial r}},\label{eq:ch}\\
\nu_{T} & = & \mathrm{Pr}_{T}\frac{u_{c}\ell_{c}}{3},\label{eq:nu}\\
\eta_{T} & = & \mathrm{Pm_{T}^{-1}}\frac{u_{c}\ell_{c}}{3},\nonumber \\
u_{c} & = & \frac{\ell_{c}}{2}\sqrt{-\frac{g}{2c_{p}}\frac{\partial\overline{s}}{\partial r}}\left(1+\mathrm{erf}\left(-\frac{r}{d}\left(\frac{r_{b}}{r}+0.01\right)\right)\right),
\end{eqnarray}
where we take into account the smooth decrease of $u_{c}$ toward
the bottom of the convection zone; here, we put $d=0.025R$. The dynamo
domain is divided for two parts. The convection zone extends from
$r_{b}=0.728R$ to $r_{e}=0.99R$. The model employs the overshoot
layer below the convection zone to describe the transition from the
differential rotation of the convection zone to the rigid rotation
in the radiative zone at the position $r_{i}=0.67R$. The given depth
of the overshoot layer allows us to bring the parameters of the radial
gradient of the angular velocity close to the helioseismology results.
In the transition region the turbulent parameters are defined following
the results of \citet{Ludwig1999}, also see \citet{Paxton2011}.\textbf{
}We apply an exponential decrease of all turbulent coefficients (except
the eddy viscosity and eddy diffusivity) with decrement -100, i.e.,
they are multiplied by a factor of $\exp\left(-100z/R\right)$, where
$z$ is the distance from the bottom of the convection zone. The eddy
viscosity and eddy diffusivity are kept finite at the bottom of the
tachocline, i.e., for the eddy viscosity coefficient profile within
the tachocline we put\textbf{ 
\begin{equation}
\nu_{T}^{(t)}=\frac{\nu_{T}^{(c)}}{\left(\nu_{T}^{(0)}+\nu_{T}^{(c)}\right)}\left(\nu_{T}^{(0)}+\nu_{T}^{(c)}\exp\left(-100z\right)\right),\label{nut}
\end{equation}
}where $\nu_{T}^{(c)}$ is the value at the bottom of the convection
zone, $\nu_{T}^{(0)}$ is the value inside the tachocline, $z$ is
the distance from the bottom of the convection zone. We use $\nu_{T}^{(c)}/\nu_{T}^{(0)}=20$
in the model, and we do similarly for the magnetic eddy diffusivity.
The angular velocity profile, $\Omega\left(r,\theta\right)$, and
the meridional circulation, $\mathbf{\overline{U}}^{(m)}$, are defined
by conservation of the angular momentum and azimuthal vorticity $\overline{\omega}=\left(\mathbf{\nabla}\times\mathbf{\overline{U}}^{(m)}\right)$.
In this paper we use the kinematic models excluding the magnetic feedback
on the large-scale flow and heat transport. The model shows an agreement
of the angular velocity profile with helioseismology results for $\mathrm{Pr}_{T}=3/4$.
The dynamo models with local $\boldsymbol{\overline{\mathcal{E}}}$
show cycle period of $22$ years when $\mathrm{Pm}_{T}=10$ and $C_{\alpha}=0.04$2, see futher details in \cite{PK19}.

\begin{figure*}
\includegraphics[width=0.95\textwidth]{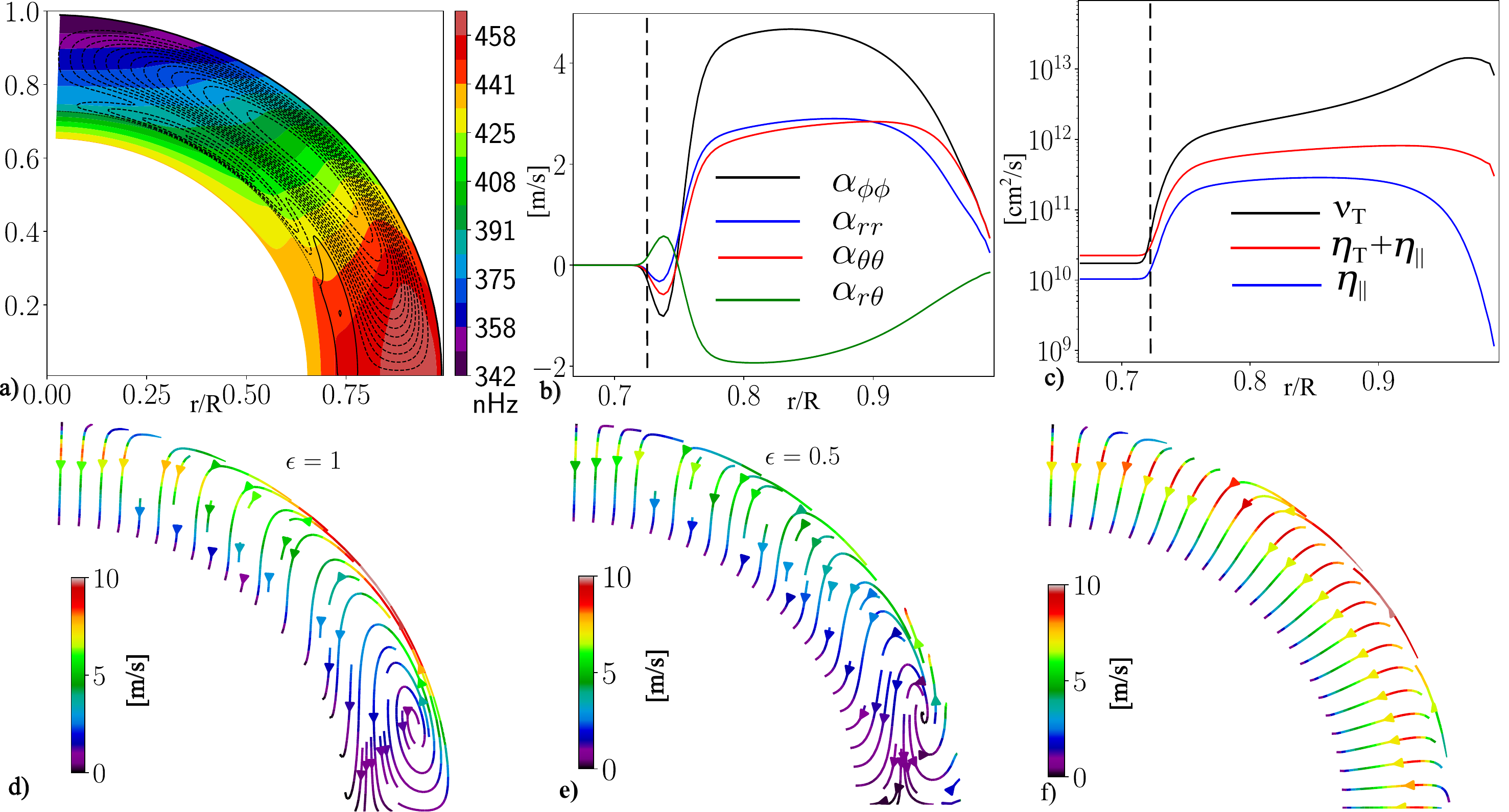} \caption{ a) The meridional circulation (streamlines) and the angular velocity
distributions; the magnitude of circulation velocity is of 13 m/s
on the surface at the latitude of 45$^{\circ}$; b) the $\alpha$-effect
tensor distributions at the latitude of 45$^{\circ}$, the dash line
shows the convection zone boundary; b) radial dependencies of the
total, $\eta_{T}+\eta_{||}$, and the rotational induced part, $\eta_{||}$,
of the eddy magnetic diffusivity, the eddy viscosity profile, $\nu_{T}$;
d) streamlines of the effective drift velocity of large-scale toroidal
magnetic field due to the turbulent pumping and the meridional circulation
for the case of equipartition between the intensity of the magnetic
fluctuations and turbulent motions, $\epsilon=1$; e) the same as
d) when the energy of the magnetic fluctuations is less by factor
two than the energy of turbulent flows; f) streamlines of the effective
drift velocity of large-scale poloidal magnetic field. We use \textsc{numpy/scipy}
\citep{harris2020array,2020SciPyNMeth} together with \textsc{matplotlib
\citep{Hunter2007} }and\textsc{ pyvista \citep{sullivan2019pyvista}
}for post-processing and visualization.}\label{fig1}
\end{figure*}

Figure \ref{fig1} shows the profiles of the large-scale flows, the
hydrodynamic $\alpha$ effect and the diffusivity profiles. We note
the inverse sign of the $\alpha$ effect tensor components and the
rotational quenching of the turbulent diffusivity profile toward the
bottom of the convection zone which is marked by the dashed line.
Figure \ref{fig1}d), e) and f) show streamlines of the effective
drift velocity of large-scale  magnetic field due to the turbulent
pumping and the meridional circulation. We see that the effective
drift patterns are different for the toroidal and poloidal magnetic
fields. Also, it depends on the magnetic fluctuations via the equipartition
parameter $\epsilon=\left\langle b^{(0)2}\right\rangle /4\pi\overline{\rho}\left\langle u^{(0)2}\right\rangle $
because of the diamagnetic pumping. These effects were extensively
discussed in \cite{Krivodubskij1987,Kitchatinov1991}.

In the paper we employ the conservation of the total magnetic helicity
\citep{Hubbard2012,Pipin2013c}, $\mathcal{H}_{V}=\int\left(\left\langle \boldsymbol{a\cdot}\boldsymbol{b}\right\rangle +\boldsymbol{\overline{A}}\cdot\overline{\boldsymbol{B}}\right)\mathrm{d}V$,
where integration is done over the volume of the convection zone.
The differential form of the conservation law is as follows
\begin{eqnarray}
\left(\frac{\partial}{\partial t}+\left(\boldsymbol{\overline{U}}\cdot\nabla\right)\right)\left(\left\langle \boldsymbol{a\cdot}\boldsymbol{b}\right\rangle +\boldsymbol{\overline{A}}\cdot\overline{\boldsymbol{B}}\right) & = & -\frac{\left\langle \boldsymbol{a\cdot}\boldsymbol{b}\right\rangle }{R_{m}\tau_{c}}-\nabla\cdot\boldsymbol{F}^{\left\langle ab\right\rangle }.\label{eq:helcon}
\end{eqnarray}
where $\boldsymbol{F}^{\left\langle ab\right\rangle }$ is the turbulent
flux of the magnetic helicity density, $\left\langle \boldsymbol{a\cdot}\boldsymbol{b}\right\rangle $.
In this equation, we neglect effects of the Ohmic diffusion of the
small- and large-scale magnetic helicity, and the Ohmic diffusion
of the large-scale current helicity. The Ohmic dissipation of the
small-scale current helicity was approximated as follows \citep{Kleeorin1999}
:
\[
\eta\left\langle \boldsymbol{b}\cdot\boldsymbol{j}\right\rangle \approx\frac{\left\langle \boldsymbol{a\cdot}\boldsymbol{b}\right\rangle }{R_{m}\tau_{c}},
\]
 where we put $R_{m}=10^{6}$. Following the results of \citet{Subramanian2023ApJ}
(hereafter, GS2023) we take $\boldsymbol{F}^{\left\langle ab\right\rangle }$
in the following form, 
\begin{equation}
\boldsymbol{F}^{\left\langle ab\right\rangle }=\boldsymbol{F}_{\mathrm{D}}+\boldsymbol{F}_{\mathrm{RA}}+\boldsymbol{F}_{\mathrm{NV}},\label{eq:fluxh}
\end{equation}
where $\boldsymbol{F}_{\mathrm{D}}$ is the Fickian flux by the turbulent
diffusion of the magnetic helicity; the $\boldsymbol{F}_{\mathrm{RA}}$
and $\boldsymbol{F}_{\mathrm{NV}}$ stand for the random advection
flux and the new Vishniac flux (NV), respectively. For $\boldsymbol{F}_{\mathrm{D}}$ GS2023
got
\begin{equation}
\boldsymbol{F}_{\mathrm{D}}=-\frac{7}{27}\left(1+\epsilon\right)u_{c}\ell_{c}\boldsymbol{\nabla}\left\langle \boldsymbol{a\cdot}\boldsymbol{b}\right\rangle .\label{eq:FD}
\end{equation}
Calculations of GS2023 does not take into account the effect of the
Coriolis force and the large-scale magnetic field on the helicity
flux. Results of P08 and KR22 show a complicated structure of the
helicity fluxes under effects of  the global rotation and large-scale magnetic field. In
this study we take into account the rotational and magnetic quenching
of the transport effects in heuristic way, in following the results of
P08 and using the same quenching functions as for the magnetic part
of the alpha effect. These quenching functions are needed for the
numerical stability of the model. Following to results of P08, the
effects of the magnetic helicity generation in the rotating turbulence
are subjected to the quenching by means of the Coriolis force. Since
the diffusive flux, $\boldsymbol{F}_{\mathrm{D}}$ is proportional
to the magnetic eddy diffusivity we apply the same rotational quenching
functions as for the isotropic part of the eddy diffusivity tensor,
i.e, we put 
\begin{equation}
\boldsymbol{F}_{\mathrm{D}}=-C_{D}\eta_{T}\left(1+\epsilon\right)\left(2f_{1}^{(a)}-f_{2}^{(d)}\right)\boldsymbol{\nabla}\left\langle \boldsymbol{a\cdot}\boldsymbol{b}\right\rangle ,\label{eq:FDq}
\end{equation}
where we put $\eta_{T}=\mathrm{Pm_{T}^{-1}}\frac{u_{c}\ell_{c}}{3}$
and for compatibility with the results of \citet{Mitra2010} we will
use $C_{D}\le1$ (the range 0.1-0.5 is employed in the dynamo model).
Following to the results of GS23 we get for the random advection flux,$\boldsymbol{F}_{\mathrm{RA}}$,
\begin{eqnarray}
F_{\mathrm{RA}} & = & u_{c}^{2}\tau_{c}f_{2}^{(a)}\left(\frac{1}{18}+\frac{7\epsilon}{27}\right)\boldsymbol{\nabla}\log\overline{\rho}u_{c}^{2}\left\langle \boldsymbol{a\cdot}\boldsymbol{b}\right\rangle \label{eq:Fra}\\
 & + & F_{\mathrm{RA}}^{W},\label{eq:Bu0}\\
F_{\mathrm{RA}}^{W} & = & -\frac{13\tau_{c}^{2}}{135}f_{2}^{(a)}\frac{\left\langle \boldsymbol{a\cdot}\boldsymbol{b}\right\rangle ^{2}}{4\pi\overline{\rho}\left\langle u^{2}\right\rangle }\boldsymbol{\overline{W}}\label{eq:Bu}
\end{eqnarray}
where, $\overline{\boldsymbol{W}}=\boldsymbol{\nabla}\times\overline{\boldsymbol{U}}$;
also, in transforming the expression given by GS23 we put $\left\langle b^{(0)2}\right\rangle =4\pi\epsilon\overline{\rho}u_{c}^{2}$
and $\left\langle b^{(0)}\cdot\nabla\times b^{(0)}\right\rangle =\left\langle \boldsymbol{a\cdot}\boldsymbol{b}\right\rangle /4\pi\overline{\rho}\ell_{c}^{2}$.
Using $\boldsymbol{\nabla}\cdot\overline{\boldsymbol{W}}=0$, we transform
$-\boldsymbol{\nabla}\cdot F_{\mathrm{RA}}^{W}$ to 
\begin{equation}
-\boldsymbol{\nabla}\cdot F_{\mathrm{RA}}^{W}=\frac{13}{135}\boldsymbol{\overline{W}}\cdot\nabla\left(\tau_{c}^{2}f_{2}^{(a)}\frac{\left\langle \boldsymbol{a\cdot}\boldsymbol{b}\right\rangle ^{2}}{4\pi\overline{\rho}\left\langle u^{2}\right\rangle }\right).\label{eq:Bun}
\end{equation}
We do the same reduction for GS23's $\boldsymbol{F}_{\mathrm{NV}}$
flux. After such simplification we get 

\begin{eqnarray}
-\boldsymbol{\nabla}\cdot F_{\mathrm{NV}} & = & 4\pi\epsilon C_{\mathrm{NV}}C_{2}\boldsymbol{\nabla}\left(\overline{\rho}\tau_{c}^{2}f_{2}^{(a)}\psi_{\alpha}\left(\beta\right)\right)\cdot\boldsymbol{\overline{W}},\label{eq:Fnv}
\end{eqnarray}
 The constant in the last equation is $C_{2}=203/5400$ and it is
$1/12$ if we consider the results of KR22 (see, expression $\boldsymbol{F}^{(S0)}$
in their Eq(E6)). The free parameter $C_{\mathrm{NV}}\ge1$ measures
the efficiency of the helicity production. In our reduction of the
$\boldsymbol{F}_{\mathrm{NV}}$ given by GS23 we neglect the combined
term with the numerical factor $C_{1}+C_{2}+C_{3}+C_{4}\approx3\cdot10^{-4}$
which is two order of magnitude less than that given by Eq(\ref{eq:Fnv}).
For the numerical stability we include the algebraic quenching of
the flux given by the same function as for the $\alpha$ - effect.
\begin{figure}
\includegraphics[width=0.9\linewidth]{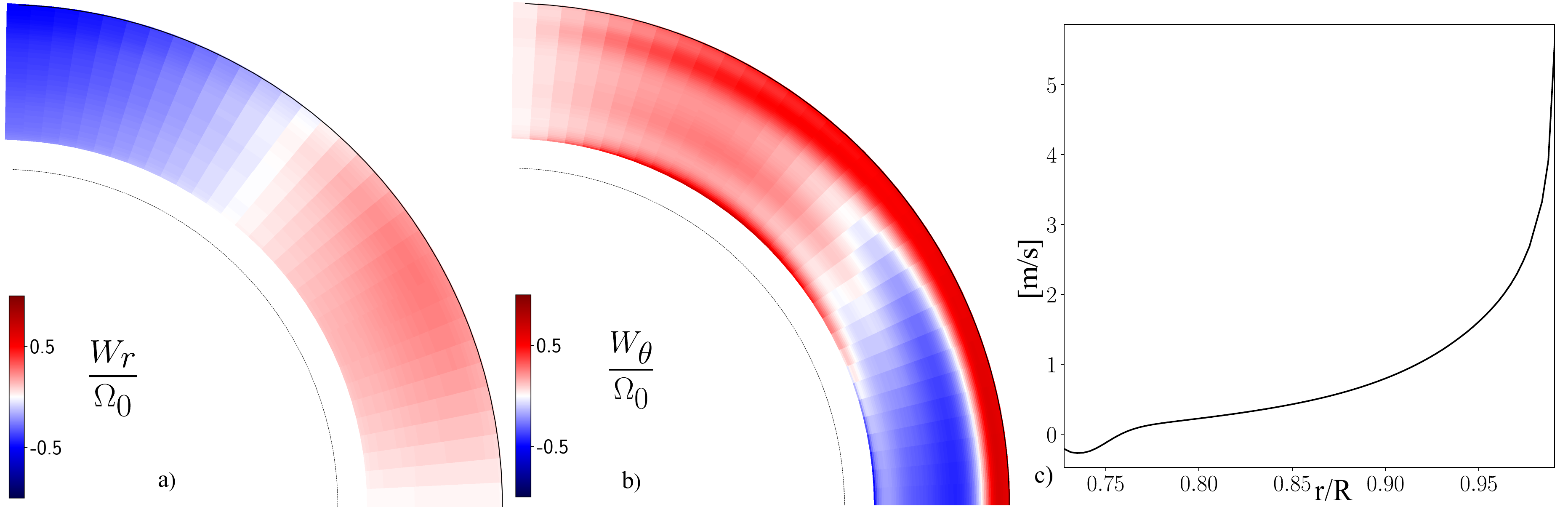}

\caption{Panels a) and b) show the components large-scale vorticity as due
to the differential rotation of the Sun. It is noteworthy that $W_{r}$
and $W_{\theta}$ are antisymmetric and symmetric about the equator,
respectively; c) the small-scale helicity transport velocity by $-\nabla\cdot F_{\mathrm{RA}}$.}\label{fig2}
\end{figure}

The large-scale vorticity is computed from the angular velocity distribution.
In the spherical coordinate system we have $\boldsymbol{\overline{W}}={\displaystyle \left(\sin\theta\frac{\partial\Omega}{\partial\theta}+2\cos\theta\delta\Omega\right)\mathbf{e}^{r}}-\sin\theta{\displaystyle \left(r\frac{\partial\Omega}{\partial r}+2\delta\Omega\right)\mathbf{e}^{\theta}}$,
where  $\delta\Omega=\Omega-\Omega_{0}$ and  $\Omega_{0}$ is the
global angular velocity of the star. We neglect the azimuthal part
of the large-scale vorticity due to the meridional circulation. Figure
\ref{fig2} shows the distribution of the large-scale vorticity in
the solar convection zone and the velocity of the small-scale helicity
transport by $-\nabla\cdot F_{\mathrm{RA}}$. The latter is outward
in the main part of the convection zone and it is inward near the
bottom.

Both the flux $\boldsymbol{F}_{\mathrm{NV}}$ and $F_{\mathrm{RA}}^{W}$
describe the magnetic helicity generation from the large-scale flow
and the small-scale magnetic fluctuations due to turbulent dynamo.
Hence they can produce the large-scale dynamo effect even if the kinetic
$\alpha$ effect is negligible. Also, we see that the large-scale
vorticity affect the nonlinear balance of the helicity distributions
in the radial and latitudinal directions. 

The dynamo model employs the zero magnetic field boundary conditions
at the bottom of the overshoot layer, $r_{i}=0.67R$, at the top we
smoothly connect the dynamo solution with the outer harmonic magnetic
field \citep{Bonanno2016}, that satisfies, 
\begin{equation}
\left(\nabla^{2}+k^{2}\right)\overline{\boldsymbol{B}}^{(e)}=0,\label{eq:harm}
\end{equation}
for the region $r_{t}<r<2.5R$ and the radial magnetic field for $r\ge2.5R$.
We use $kR=0.1$ as suggested by the results of the above-cited paper.
To connect the external magnetic field with the dynamo region we employ
the continuity of the normal component of the magnetic field and the
tangential component of mean electromotive force, i.e., we put
\begin{equation}
\eta_{T}\left(\boldsymbol{\nabla}\times\overline{\boldsymbol{B}}\right)_{\theta}=\eta_{T}^{+}\left(\boldsymbol{\nabla}\times\overline{\boldsymbol{B}}^{(e)}\right)_{\theta},\label{eqBC}
\end{equation}
where $\eta_{T}$ is the surface value of the turbulent magnetic diffusivity
and $\eta_{T}^{+}$ is the mean effective diffusivity in the stellar
corona. The solar type dynamo can be obtained when $\eta_{T}^{+}/\eta_{T}\gg$1,
i.e, when the corona is close to the ideal insulator state. Following
the results of \citet{Pipin2025ApJ} we put $\eta_{T}^{+}/\eta_{T}=1000$.
The numerical code employs the pseudo-spectral method of integration
in latitude with 64 Legendre nodes from the North to the South poles.
In the radius it utilizes the finite differences with 120 mesh points
from the bottom of the dynamo domain $r_{i}=0.67R$ to the top $r=0.99R$.
The Table \ref{tab1} contains the description of the runs which will
be discussed in the next chapter. Each run was started from the weak
poloidal field of the equally mixed parity and 1 G strength. The solution
was followed until it reach the stationary stage. The models with
$\epsilon<1$ have a higher dynamo threshold than the case $\epsilon=1$. 

\begin{table}
\centering{}\caption{The models and parameters. Here, $B_{\phi}/\Phi$ marks the magnitude
and the total unsigned of the toroidal magnetic field in the dynamo
region; $B_{r}$ shows the strength of the polar magnetic field at
the surface; $P_{\mathrm{dyn}}$ is the full magnetic cycle period;
the Parity parameter shows the type of the equatorial symmetry of
the magnetic fields, where D stands for the dipole-type symmetry magnetic
field, Q stands for the quadrupole-type and D+ wQ is mix of the dipole
type symmetry with a weak quadrupole-type symmetry, respectively.
}\label{tab1}
\begin{tabular}{ccccccccccc}
\hline 
Model & $C_{\alpha}$ & $\epsilon$ & $C_{\mathrm{NV}}$ & $F_{\mathrm{RA}}$ & $C_{D}$ & $B_{\phi}${[}kG{]} & $\Phi$$10^{24}${[}Mx{]} & $B_{r}$,{[}G{]} & $P_{\mathrm{dyn}}$,{[}yr{]} & Parity\tabularnewline
\hline 
M0 & 0.045 & 1 & 0 & 0 & 0.1 & 4.1 & 1.45 & 30 & 20.8 & D\tabularnewline
\hline 
M1 & 0.055 & 0.5 & 0 & 0 & 0.1 & 8.1 & 1.25 & 53 & 26.5 & D+ w Q\tabularnewline
\hline 
M2 & 0.045 & 1 & 1 & 1 & 0.3 & 5.8 & 2.1 & 60 & 23.4 & D+ w Q\tabularnewline
\hline 
M3 & 0.055 & 0.5 & 1 & 1 & 0.3 & 8.3 & 1.02 & 58 & 24.5 & D+ w Q\tabularnewline
\hline 
\end{tabular}
\end{table}

\section{Results}

\begin{figure}
\includegraphics[width=0.9\linewidth]{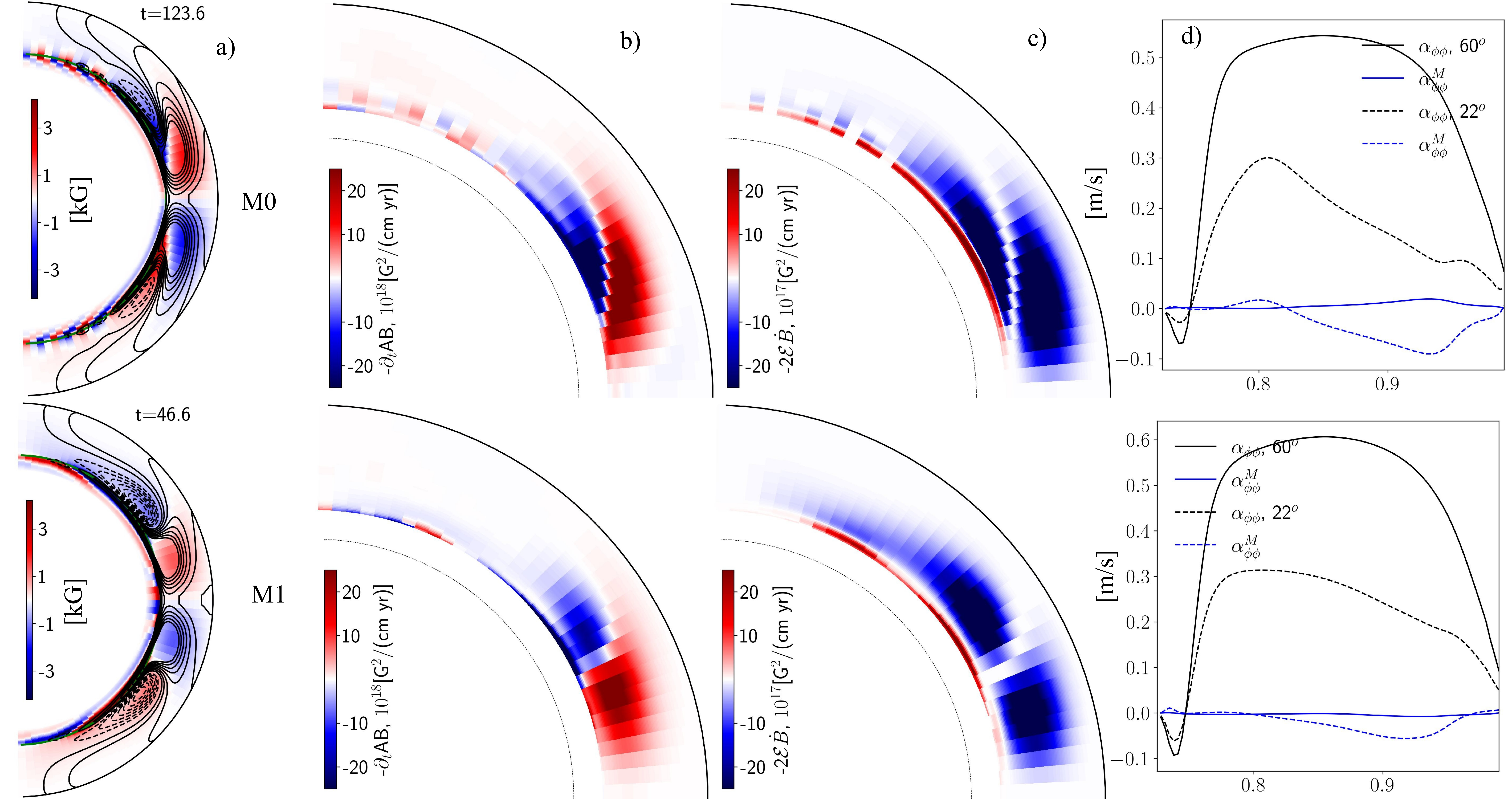}

\caption{Snapshots of the models M0 and M1 for the decaying phase of the magnetic
cycle: a) the large-scale magnetic field distribution, color image
shows the toroidal magnetic field and the streamlines show the poloidal
magnetic field; b) the small-scale magnetic helicity density generation
rate by the large-scale dynamo ; c) the same as b) for the $\alpha$
effect contribution; d) the radial profiles of the $\alpha$ effect,
$\alpha_{\phi\phi}^{M}$ shows the contribution of the small-scale
magnetic helicity. }\label{fig3}

\end{figure}

Figure \ref{fig3} show the snapshots of the reference models M0 and
M1 for the decaying phase of the magnetic cycle. We see that the reduction
of the intensity of the magnetic fluctuations, i.e., $\epsilon<1$
results to a considerable change in the pattern of the dynamo activity.
Firstly, the radial propagation of the dynamo waves is not seen in
the case M1. Secondly, the activity of the toroidal filed is concentrated
strongly toward the bottom of the convection zone. This effect result
to a decrease of the magnitude of the toroidal magnetic field flux
in the dynamo in the run M1 in compare to the run M0. The run M1 show
the longer dynamo period than the run M0 because the toroidal magnetic
field evolves in the region with low magnetic diffusivity. All these
effects in the run M1 are mainly because of the diamagnetic pumping
pumping which is not active in the case $\epsilon=1$. The d) column
in Figure \ref{fig3} show variations of the total $\text{\ensuremath{\alpha}}$
effect, and its magnetic part which is due to the small-scale magnetic
helicity generation. The dynamic variations of the $\text{\ensuremath{\alpha}}$
effect at low latitudes in the run M0 are greater than in the run
M1.

\begin{figure}
\includegraphics[width=0.9\linewidth]{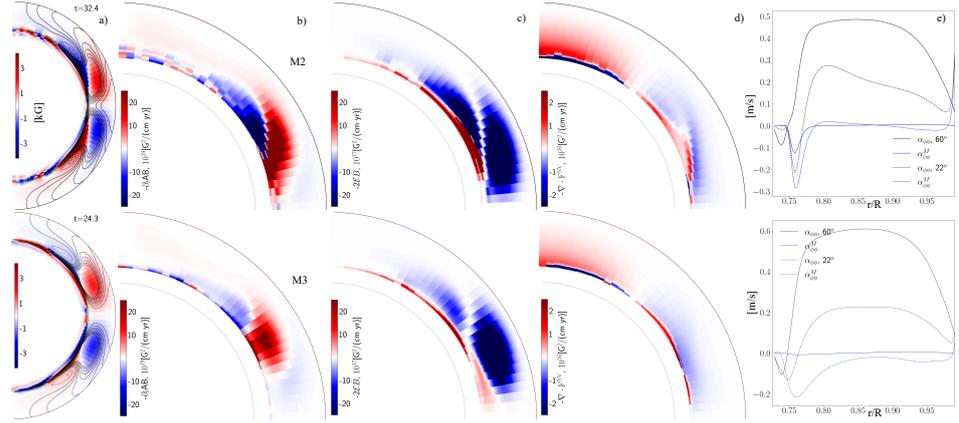}

\caption{The same as \ref{fig3} for the models M2 and M3. Here, the column
d) show the new Vishniac flux contribution, $-\boldsymbol{\nabla}\cdot F_{\mathrm{NV}}$.
Note the substantial changes of the alpha effect near the bottom of
the convection zone and at high latitude near the surface. Animations
of this figure show variations of the magnetic field and the helicity
generation rate distributions, as well as variations of the $\alpha$
effect profiles during 3 magnetic cycles of the runs M2 and M3.}\label{fig4}
\end{figure}

Addition of the helicity generation due the large-scale vorticity
results to a strong modification of the total $\alpha$ effect both
near the bottom and the top of the convection zone. Figure \ref{fig4}
shows the snapshots of the magnetic field distribution and helicity
generation rate in the dynamo model for the decaying phase of the
activity cycle. The runs show that in the main part of the convection
zone the small-scale helicity generation rate by the source $-\boldsymbol{\nabla}\cdot F_{\mathrm{NV}}$
is an order of magnitude less than that by the standard $\alpha$
effect. However the effects have same magnitudes near the bottom of
the convection zone and near the surface in the polar regions of the
star. The run M2 demonstrates the significant change of the total
$\alpha$ effect in these regions. The animations of constructed using
the results of run M2 and M3 show significant variations of the $\alpha$
effect in the magnetic cycle due to the new source of the magnetic
helicity generation. Similarly to the case M1, the run M3 shows a
decrease of the $\alpha$ effect variations in the bulk of the convection
zone in compare to the run M2. The increase of the total $\alpha$
effect near the surface in the polar regions the run M2 in compare
to run M0 results to a stronger magnitude of the magnetic cycle. Simultaneously,
run M2 shows a wider region of the inverted $\alpha$ effect at low
latitudes near the bottom of the convection zone that the run M0.
Following to the Parker-Yoshimura rule this amplifies the propagation
of the dynamo wave \emph{in this region} toward the equator and bottom-ward.
This seems affect the period of the dynamo cycle in the bulk of the
convection zone, see Table 1.

Similarly to the runs M0 and M1 a decrease the level of the magnetic
fluctuations to $\epsilon=1/2$ results to the change of the dynamo
wave propagation pattern and the increase of the toroidal magnetic
field near the bottom of the convection zone and in overshoot layer.
The latter results to a decrease of the $\alpha$ effect modulation
near the bottom of the convection zone. The run M3 shows smaller modulation
of the $\alpha$ effect near the top of the convection zone in compare
to the run M2.

\begin{figure}
\includegraphics[width=0.99\linewidth]{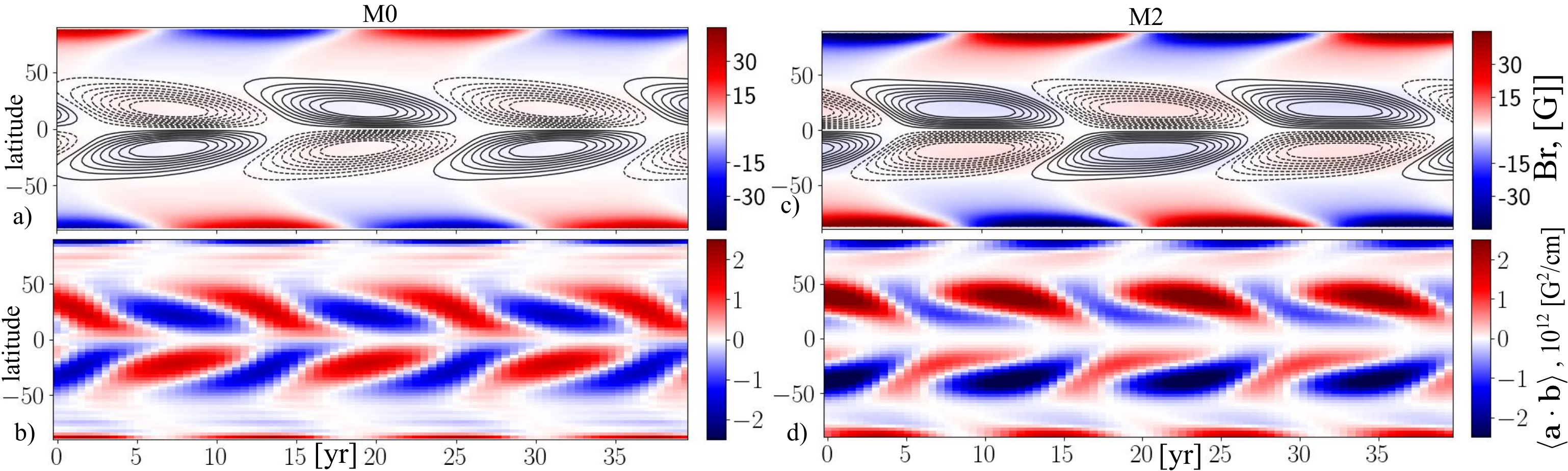}

\caption{a) The time latitude diagrams of the large-scale magnetic field in
the model M0, contours show the toroidal magnetic field at $r=0.9R$;
b) the time latitude diagram of the small-scale helicity density evolution
at the surface; c) and d) show the same for the run M2.}\label{fig5}
\end{figure}

Figure \ref{fig5} illustrates the increase of the dynamo efficiency
in the run M2 in compare to the run M0 by the time-latitude diagrams
of the large-scale magnetic field and small-scale magnetic helicity.
In the northern hemisphere of the star the run M2 shows an increase
generation of the positive small-scale magnetic helicity around 50$^{\circ}$
latitude during the decaying phase of the magnetic cycle when the
radial magnetic field grows in the polar regions. The Table 1 shows
the magnitude of the polar magnetic field in the run M2 is by factor
two larger than in the run M0.

In this paper I discuss only a few runs with the purpose of illustration
of the dynamo effects due to the large-scale vorticity and the fluctuating
magnetic field that stem from the small-scale dynamo. Let's discuss
the the effect of the parameter variations. When the small-scale dynamo
absent, $\epsilon\ll$1, then the helicity transport by the turbulent
motions stratification effects decrease by the factor of magnitude,
see Figure \ref{fig2}c. The flux $\text{-\ensuremath{\boldsymbol{\nabla}\cdot F_{\mathrm{NV}}}}$
disappears in this case. The moderate reduction of the small-scale
parameter $\epsilon$, see the runs M1 and M3 results into decrease
of the dynamo efficiency of the convective envelope. The decrease
of $\epsilon$ results into increase of the critical dynamo threshold
of the $\alpha$ effect. We find the longer dynamo period with the
increase of the turbulent diffusion flux. It results into reduction
the magnetic cycles overlap, as well. The latter affect the extended
mode of the torsional osculations \citep{PK19}. We find that the
new dynamo source due to $-\boldsymbol{\nabla}\cdot F_{\mathrm{NV}}$
can produce the large-scale dynamo even in the case when the kinetic
$\alpha$ effect is zero. We postpone discussion of this case as well as effect of the nonlinear helicity transport  to a
separate paper.

\section{Discussion and Conclusions}

In this paper we discussed the large-scale dynamo effects of the divergent
helicity flux due to the large-scale vorticity and small-scale dynamo
magnetic fluctuations. The magnetic helicity flux is usually considered
as a solution of the so-called $\alpha$ -effect catastrophic quenching
\citep{Vainshtein1992}. In this case the helicity generation follows
conservation of the total magnetic helicity in the large-scale dynamo
as a response for generation large-scale magnetic field by the kinetic
$\alpha$ effect. Results of \citet{Kleeorin2000,Vishniac2001,Field2002}
showed the catastrophic quenching can be alleviated if the magnetic
helicity transport by the turbulent motions is taken into account.
Generation of the small-scale magnetic helicity in the $\alpha^{2}\Omega$
dynamos is a nonlinear about the $\overline{\boldsymbol{B}}$- field
effect. The analytical results of \citet{Kleeorin2022,Subramanian2004}
showed that the small-scale helicity can be generated by the large-scale
vorticity and the chaotic small-scale magnetic field that stem from
the small-scale dynamo. Generation of the preferred hemispheric helicity
from the random magnetic field distribution by the differential rotation
is known for a long time, see e.g., \citet{Berger1984,Berger2000,PariDemBer2005AA}.
The results of the numerical model show that the differential rotation
can generate the negative helicity in the near equatorial regions
at the northern hemisphere and positive at the southern \citep{Prior2014ApJ,Pipin2025ApJ}.
The opposite pattern is generated at high latitudes. The generated
magnetic helicity participates in the dynamo process and it must be
taken into account. The results of KR22 and GS23 formulate this effect
in a general analytical form. Here, for the first time we consider
this effect  for the solar type dynamos. The new dynamo source in
the studied dynamo model ss constructed in heuristic way because we
do not know the impact of the Coriolis force and the large-scale magnetic
field on $F_{\mathrm{NV}}$. This affect all our conclusions constructed
on the base of the presented runs.

The results of our runs shows that the new source of the magnetic
helicity modifies the patterns of the dynamo waves propagation inside
the convection zone, increases the amplitude and the dynamo period.
These changes are caused by modification of the $\alpha$ effect.
The source $-\boldsymbol{\nabla}\cdot F_{\mathrm{NV}}$ is concentrated
toward the boundaries of the convection zone because of variation
of the differential rotation, the convective turnover time $\tau_{c}$
and the mean density stratification. These factors results to a modulation
of the $\alpha$ effect near the boundaries. Near the bottom the source
$-\boldsymbol{\nabla}\cdot F_{\mathrm{NV}}$ increases the inversion
of the $\alpha$ effect. In following the Parker-Yoshimura rule \citep{Yoshimura1975},
the negative $\alpha$ ( in the northern hemisphere) and the positive
gradient of the angular velocity result into amplification of the
equatorward dynamo waves in equatorial regions near the bottom of
the convection zone. Simultaneously the negative shear in the polar
regions together with the negative gradient of the mean density and
the small-scale dynamo magnetic field amplify generation of the poloidal
magnetic field near the surface in the polar regions of the star.
The latter means that the polar magnetic field can be regenerated
in-situ if there is the toroidal magnetic field there. Therefore the
large-scale vorticity amplifies the action of the kinetic $\alpha$
-effect in the polar regions of the star. Similarly to the $\alpha^{2}\Omega$
dynamo, see \citet{Kitchatinov2011b}, the new dynamo source $-\boldsymbol{\nabla}\cdot F_{\mathrm{NV}}$
is likely to be important for the fast rotating stars that have the
finite amplitude of the differential rotation.

Let us summarize our findings. Using the mean-field dynamo model we
study the effects of the large-scale vorticity and the magnetic fluctuations
on the large-scale dynamo. The intensity of the magnetic fluctuations
is given by the equipartition parameter $\epsilon=\left\langle b^{(0)2}\right\rangle /4\pi\overline{\rho}\left\langle u^{(0)2}\right\rangle \le1$.
Its effect on the small-scale helicity production was estimated earlier
in analytical studies of \citet{Pipin2008a,Kleeorin2022,Subramanian2023ApJ}
using the mean-field MHD framework. Our dynamo models show a crucial
role of the magnetic helicity production via the large-scale vorticity
in the large-scale dynamo process. It affects the magnetic field generation
near the boundaries of the dynamo domain because of the strong variation
of the mean turbulent parameters in these regions. We hope that the
future study will clarify the nonlinear impact of the global rotation
and magnetic fields on the level of the magnetic helicity production
by the differential rotation and the magnetic fluctuations.
\reftitle{References}
\bibliography{dyn.bib}

\dataavailability{The data of the models are available by request from the author.} 

\acknowledgments{The author thanks the Ministry of Science and Higher Education of
the Russian Federation for financial support (the project No.0278-2026-0001)}

\conflictsofinterest{The authors declare no conflicts of interest.} 

\subsection*{Appendix A.}

\paragraph{The $\alpha$ effect.}

In following to \citet{Pipin2008a} (hereafter P08) we have
\begin{eqnarray*}
\alpha_{ij}^{{\rm K}} & = & \delta_{ij}\left\{ \left(\boldsymbol{e}\cdot\boldsymbol{\Lambda}^{(\rho)}\right)f_{10}^{(a)}+\left(\boldsymbol{e}\cdot\boldsymbol{\Lambda}^{(\eta)}\right)f_{11}^{(a)}\right\} \\
 & + & e_{i}e_{j}\left\{ \left(\boldsymbol{e}\cdot\boldsymbol{\Lambda}^{(\rho)}\right)f_{5}^{(a)}+\left(\boldsymbol{e}\cdot\boldsymbol{\Lambda}^{(\eta)}\right)f_{4}^{(a)}\right\} \\
 & + & \left(e_{i}\Lambda_{j}^{(\rho)}+e_{j}\Lambda_{i}^{(\rho)}\right)f_{6}^{(a)}+\left(e_{i}\Lambda_{j}^{(\eta)}+e_{j}\Lambda_{i}^{(\eta)}\right)f_{8}^{(a)},
\end{eqnarray*}
and 
\[
\alpha_{ij}^{{\rm M}}=\delta_{ij}f_{2}^{(a)}-e_{i}e_{j}f_{1}^{(a)},
\]
where the functions $f_{n}^{(a)}$ , $f_{6.7}^{'(a)}$ define the
Coriolis number dependence and dependence of the alpha effect tensors
on the relative intensity of the magnetic fluctuations $\epsilon=\left\langle b^{(0)2}\right\rangle /4\pi\overline{\rho}\left\langle u^{(0)2}\right\rangle \le1$
that stems from the small-scale dynamo. The latter is important for
the NV helicity flux (see, GS2023). In the previous studies we simply
use $\epsilon=1$. Here we consider this as another parameter of the
dynamo model. We put all the quenching functions in Appendix. The
stratification scales parameters are $\mathbf{\boldsymbol{\Lambda}}^{(\rho)}=\ell_{c}\boldsymbol{\nabla}\log\overline{\rho}$
and $\mathbf{\boldsymbol{\Lambda}}^{(\eta)}=\ell_{c}\boldsymbol{\nabla}\log\tau_{c}\left\langle u^{2}\right\rangle $.
It is worth to note that for the nearly adiabatic stratification of
the stellar convection zone we have $\Lambda^{(\rho)}=-\alpha_{\mathrm{MLT}}/\gamma$,
where $\gamma$ is the adiabatic index and $\alpha_{\mathrm{MLT}}$
is the mixing-length parameter we use $\ell_{c}=\alpha_{\mathrm{MLT}}H_{p}$
and $H_{p}=\left|\nabla\log\overline{P}\right|$, $\overline{P}$
is the mean pressure profile. The $\mathbf{\boldsymbol{\Lambda}}^{(\eta)}$
changes from about $\approx-0.5$ in the main part of the convection
zone and it goes to $\gg1$ below $0.73R$ near the bottom. Therefore,
the $\alpha_{ij}^{{\rm K}}$ inverses the sign near the bottom of
the convection zone. Functions $f_{n}^{(a)}\left(\Omega^{*}\right)$
were defined by \citet{Pipin2008a}, $\Omega^{*}=2\tau_{c}\Omega_{0}$,
and $\Omega_{0}/2\pi=432$nHz.

\begin{eqnarray*}
f_{1}^{(a)} & = & \frac{1}{4\Omega^{*\,2}}\left(\left(\Omega^{*\,2}+3\right)\frac{\arctan\Omega^{*}}{\Omega^{*}}-3\right),\\
f_{2}^{(a)} & = & \frac{1}{4\Omega^{*\,2}}\left(\left(\Omega^{*\,2}+1\right)\frac{\arctan\Omega^{*}}{\Omega^{*}}-1\right),\\
f_{4}^{(a)} & = & \frac{1}{6\Omega^{*\,3}}\left(3\left(\Omega^{*4}+6\varepsilon\Omega^{*2}+10\varepsilon-5\right)\frac{\arctan\Omega^{*}}{\Omega^{*}}\right.\\
 &  & \left.-\left((8\varepsilon+5)\Omega^{*2}+30\varepsilon-15\right)\right),\\
f_{5}^{(a)} & = & \frac{1}{3\Omega^{*\,3}}\left(3\left(\Omega^{*4}+3\varepsilon\Omega^{*2}+5(\varepsilon-1)\right)\frac{\arctan\Omega^{*}}{\Omega^{*}}\right.\\
 &  & \left.-\left((4\varepsilon+5)\Omega^{*2}+15(\varepsilon-1)\right)\right),\\
f_{6}^{(a)} & = & -\frac{1}{48\Omega^{*\,3}}\left(3\left(\left(3\varepsilon-11\right)\Omega^{*2}+5\varepsilon-21\right)\frac{\arctan\Omega^{*}}{\Omega^{*}}\right.\\
 &  & \left.-\left(4\left(\varepsilon-3\right)\Omega^{*2}+15\varepsilon-63\right)\right),\\
f_{8}^{(a)} & = & -\frac{1}{12\Omega^{*\,3}}\left(3\left(\left(3\varepsilon+1\right)\Omega^{*2}+4\varepsilon-2\right)\frac{\arctan\Omega^{*}}{\Omega^{*}}\right.\\
 &  & \left.-\left(5\left(\varepsilon+1\right)\Omega^{*2}+12\varepsilon-6\right)\right),\\
f_{10}^{(a)} & = & -\frac{1}{3\Omega^{*\,3}}\left(3\left(\Omega^{*2}+1\right)\left(\Omega^{*2}+\varepsilon-1\right)\frac{\arctan\Omega^{*}}{\Omega^{*}}\right.\\
 &  & \left.-\left(\left(2\varepsilon+1\right)\Omega^{*2}+3\varepsilon-3\right)\right),\\
f_{11}^{(a)} & = & -\frac{1}{6\Omega^{*\,3}}\left(3\left(\Omega^{*2}+1\right)\left(\Omega^{*2}+2\varepsilon-1\right)\frac{\arctan\Omega^{*}}{\Omega^{*}}\right.\\
 &  & \left.-\left(\left(4\varepsilon+1\right)\Omega^{*2}+6\varepsilon-3\right)\right).
\end{eqnarray*}

The magnetic quenching function of the hydrodynamical part of $\alpha$-effect:
\begin{equation}
\psi_{\alpha}=\frac{5}{128\beta^{4}}\left(16\beta^{2}-3-3\left(4\beta^{2}-1\right)\frac{\arctan\left(2\beta\right)}{2\beta}\right),
\end{equation}
where $\beta=|\overline{\boldsymbol{B}}|/\sqrt{4\pi\overline{\rho}u_{c}^{2}}$

\paragraph*{The turbulent pumping and eddy magnetic diffusivity.}

In the model we take into account the mean drift of large-scale field
due to the magnetic buoyancy, $\gamma_{ij}^{(\mathrm{buo})}$ , the
gradient of the mean density, $\gamma_{ij}^{(\Lambda\rho)}$ and the
diamagnetic pumping because of the turbulent intensity stratification,
$\gamma_{ij}^{(\Lambda\eta)}$: 
\begin{eqnarray}
\gamma_{ij} & = & \gamma_{ij}^{(\Lambda\rho)}+\gamma_{ij}^{(\mathrm{buo})}+\gamma_{ij}^{(\Lambda\eta)},\nonumber \\
\gamma_{ij}^{(\Lambda\rho)} & = & u_{c}f_{1}^{(a)}\left\{ \left(\boldsymbol{e}\cdot\boldsymbol{\Lambda}^{(\rho)}\right)e_{n}\varepsilon_{inj}-e_{j}\varepsilon_{inm}e_{n}\Lambda_{m}^{(\rho)}\right\} \label{eq:pump1}\\
\gamma_{ij}^{(\mathrm{buo})} & = & \frac{\alpha_{\mathrm{MLT}}u_{c}}{\gamma}\mathcal{H}\left(\beta\right)\hat{r}_{n}\varepsilon_{inj},\nonumber \\
\gamma_{ij}^{(\Lambda\eta)} & = & \left(\epsilon-1\right)u_{c}\left\{ f_{2}^{(a)}\Lambda_{n}^{(\eta)}\varepsilon_{inj}-f_{1}^{(a)}e_{j}\varepsilon_{inm}e_{n}\Lambda_{m}^{(\eta)}\right\} .
\end{eqnarray}
where the magnetic buoyancy modulation function $\mathcal{H}\left(\beta\right)$
reads
\[
\mathcal{H}\left(\beta\right)=\frac{1}{8\beta^{2}}\left(\frac{3}{\beta}\arctan\left(\beta\right)-\frac{\left(5\beta^{2}+3\right)}{\left(1+\beta^{2}\right)^{2}}\right),
\]
From Eq(\ref{eq:pump1}), we see that that $\gamma_{ij}^{(\Lambda\eta)}=0$
when $\epsilon=1$. In other words, the diamagnetic pumping disappears
when the energy of the small-scale magnetic fluctuations is in equipartition
with the energy of the convective flows. 

We employ the anisotropic diffusion tensor following the formulation
of P08 ignoring the generation effects from the large-scale current
: 
\begin{eqnarray}
\eta_{ijk} & = & 3\eta_{T}\left\{ \left(2f_{1}^{(a)}-f_{2}^{(d)}\right)\varepsilon_{ijk}+2f_{1}^{(a)}\frac{\Omega_{i}\Omega_{n}}{\Omega^{2}}\varepsilon_{jnk}\right\} \label{eq:diff-1}
\end{eqnarray}
where $f_{2}^{(d)}=f_{2}^{(d)}\left(\Omega^{*},\epsilon\right)$ is
\[
f_{2}^{(d)}=\frac{1}{4\Omega^{*\,2}}\left(\left(\left(\varepsilon-1\right)\Omega^{*\,2}+3\varepsilon+1\right)\frac{\arctan\left(\Omega^{*}\right)}{\Omega^{*}}-\left(3\varepsilon+1\right)\right)
\]

\end{document}